\documentclass[acmsmall, nonacm]{acmart}

\AtBeginDocument{%
  }

\usepackage[firstpageonly=true, color=black, pos={240pt,30pt}, angle=0, fontsize=14pt]{draftwatermark}
\SetWatermarkText{For full version of the paper accepted at \\Fourth European Workshop on Algorithmic Fairness: \\https://proceedings.mlr.press/v294/marcinc-inova25a.html}

\setcopyright{none}              
 
\begin{document}

\title{Towards Trustworthy Multi-stakeholder Recommender Systems}

\author{Katarina Marcincinova}
\affiliation{%
  \institution{Kempelen Institute of Intelligent Technologies}
  \country{Slovakia}
}
\email{name.surname@kinit.sk}

\author{Adrian Gavornik}
\affiliation{%
  \institution{Kempelen Institute of Intelligent Technologies}
  \country{Slovakia}}
\email{name.surname@kinit.sk}

\author{Matus Mesarcik}
\affiliation{%
  \institution{Kempelen Institute of Inteligent Technologies, Comenius University Bratislava}
  \country{Slovakia}}
\email{name.surname@kinit.sk}

\author{Michal Kompan}
\affiliation{%
  \institution{Kempelen Institute of Inteligent Technologies}
  \country{Slovakia}}
\email{name.surname@kinit.sk}
\renewcommand{\shortauthors}{Marcincinova K., Gavornik A., Mesarcik M., Kompan M.}

\begin{abstract}
Recommender system (RS) nowadays has to reflect, combine, and solve often contrasting requirements and expectations from dozens of stakeholders. In fact, the problem is even more complicated as each of the stakeholders can simultaneously have various objectives creating a diverse and complicated environment with several stakeholders and their objectives in which a RS operates. To address this problems the Multi-objective and Multi-stakeholder RS have been studied in the literature. However, a limited attention has been paid to trustworthiness aspects of such RS, which slowly becomes a new standard. In this paper, we highlight open challenges and important questions which need to be addressed on the way to trustworthy Multi-stakeholder RS.
\end{abstract}



\maketitle

\section{Introduction and Related Works}
Recommender systems (RSs) have tremendously impacted the online life of every user on the Web. Starting with the Goldberg’s mail recommendation system Tapestry in 1992, which aimed at routing relevant mail to competent employees~\cite{tapestry}, the field evolved to serving billions of requests for recommendations every day on social media or e-commerce platforms. The role of a regular user on the Web also evolved. The user generated content paradigm transformed a strict consumers to the role of ``prosumers'', which mixes (or switches between) two roles of a consumer and producers. Last but not least, large platforms, e.g., YouTube, Facebook, Amazon empower users to access content from many producers (e.g., businesses). This was reflected by introducing the Multi-Stakeholder (MSRS) or Multi-Objective (MORS) recommender system. While the latter does not necessarily involve different stakeholders (a user may have several objectives simultaneously), e.g., to spend a free time on social network while to educate about new concepts, this is usually not the case with multiple stakeholders (different stakeholders tend to have different, yet still multiple objectives). 
Despite the MORSs and MSRSs have been studied in the literature, a limited attention has been paid to trustworthiness of such RS. In this paper, we analyze and emphasize open challenges and important questions which need to be addressed on the way to trustworthy MSRSs.

Although RSs are typically framed as user-centric tools designed to enhance individual experiences through personalized content, these systems are increasingly recognized as complex socio-technical assemblages that affect multiple stakeholders at once~\cite{stray_building_2024,schedl_technical_2025,milano_ethical_2021,fabbri_social_2023}. Some~\cite{Boeker_2022} argue that the widespread use of RSs can have a serious impact on how humans receive information and ultimately perceive the world. Not only that, but the increasing reliance on RSs can affect our moral skills~\cite{schuster_attention_2025} and even reshape human agency as such~\cite{jylha_algorithmic_nodate}. This is important, especially in the context of very large social media platforms, where users receive information and perceive content on a daily basis, which has often been abused (e.g., for political campaigns)~\cite{partisan}. Therefore, this socio-technical understanding of RSs goes beyond the mere technological artifacts and instead acknowledges and makes explicit its entanglement with the social, cultural and political dimensions of its development and deployment. For this reason, the question of how to create ethical or trustworthy RS is becoming even more important.

In the literature, there is no single accepted definition of ``trustworthiness'' with respect to AI systems, but rather we find specific requirements that apply to it. For example, according to EGTAI~\cite{EGTAI}, ``trustworthy'' AI should be lawful, ethical, and robust. Specifically, seven key requirements for trustworthy AI systems have been formulated: human agency and oversight; technical robustness and safety; privacy and data governance; transparency; diversity, nondiscrimination and fairness; societal and environmental being; and accountability.

The discourse on trustworthy RS has mainly focused on the negative aspect or definition, i.e., by identifying and minimizing risks, such as biases, unfairness, or privacy concerns, including other harms through specific metrics and technological fixes~\cite{abdollahpouri_multistakeholder_2020,milano_ethical_2021}. The state-of-the-art MORSs and MSRSs in the context of trustworthy also usually focus on individual requirements of trustworthiness mentioned above. For example, Schedl et al.~\cite{schedl_technical_2025} identified three main dimensions of trustworthy RS - bias, fairness and non-discrimination; transparency and explainability; privacy and security. Others have focused on specific requirement, such as diversity~\cite{Shrivastava, LETARD2024123151} or fairness~\cite{Ranjbar_Kermany2021-ai, deldjoo_fairness_2024}.

It could be argued that current approaches omit the positive aspect or the definition of trustworthiness, one of which is proactive advancement of human and especially ethical values in the design of RSs as proposed by Stray~\cite{stray_building_2024}. More importantly, in the discussion about MSRSs, trustworthiness is not addressed, as there is a lack of consensus and research exploring the key implications of such systems~\cite{DELDJOO2021102662}.

We aim to address this gap by expanding the concept of trustworthy MSRSs to incorporate both negative (risk mitigation) and positive (value-sensitive) aspects. We also argue that following the tripartite structure of trustworthy AI (legal compliance, robustness, adherence to ethical norms) should be adopted to how RSs can be in alignment with fundamental rights and societal values, particularly in the European context where regulatory frameworks like the Digital Services Act (DSA)~\cite{DSA2022}, consumer-related legislation such as Unfair Commercial Practices Directive (UCPD)~\cite{UCPD2022}, and Consumer Rights Directive (CRD)~\cite{CRD2022} provide normative guidance. 
We also argue that trustworthiness comes not only from technical solutions, but from the balanced consideration of various stakeholder values, objectives, and the inevitable tensions that arise between them. 

\section{Stakeholders and Objectives}
The wide variety of stakeholders are involved and affected by the RSs - from individual users to societies as a whole. The task of identifying and conceptualizing stakeholders is therefore not trivial. As Milano et al.~\cite{milano_ethical_2021} discuss, this concerns the mismatch in ontologies between the system design and the actual stakeholders it has an impact on. They argue that this makes ethical evaluations of RSs rather challenging. This task becomes even more complex in the context of legal frameworks. There are notable overlaps between how various regulatory regimes define and categorize stakeholders. Additionally, the appropriate level of abstraction for defining stakeholders varies across contexts - from individual users, user communities, to broader societal interests. We focus primarily on RSs for social media platforms, but other areas, such as e-commerce or streaming services, are sharing similar properties. 

In the area of very large social media platforms (i.e., Instagram, Facebook, or TikTok), the most prominent stakeholders are~\cite{fabbri_social_2023, stray_building_2024, schedl_technical_2025}: end users (consumers), content producers (creators), and service providers (social media platforms). However, there are other stakeholders that are equally affected by RSs, including developers, regulators or policymakers, and even society as a whole. The ongoing discourse in the context of trustworthy RSs explores whether the items should be included as a separate stakeholder~\cite{Zheng2021}. In other words, has a recommended item a ``right'' to be recommended? We argue that it should not be considered as a separate stakeholder. However, the item's ``rights'' come naturally from the trustworthy definition and properties (i.e., diversity, novelty, or fairness).

It is obvious that the objectives of the stakeholders differ. For example, as a major stakeholder in RSs, \textbf{users (consumers)} are primarily motivated to fulfill their information needs~\cite{Schneider}. These can be further translated into various objectives, such as finding new friends, having fun in the case of social media platforms, or buying a product with the best price and required quality in the case of e-commerce~\cite{Kompan2022}. Moreover, users benefit from regulatory requirements for providers of (very large) platforms, e.g., under the DSA~\cite{DSA2022} that should also be considered as part of their objectives. These relate to transparency regarding the ranking parameters of RSs, or users should be able to modify the parameters and explanations of the main parameters. These shall at least include the most significant criteria for recommendations and reasons for the relative importance of those parameters~\cite[Article 27]{DSA2022}.

On the other hand, it is clear that different\textbf{ content producers} as another stakeholder follow their own objectives, as maximizing the amount of content consumed or increasing the number of subscribed end-users on a social media platforms; or selling out outdated products~\cite{luo2023} or increasing customer loyalty in the case of e-commerce.

Finally, as a third core stakeholder group, \textbf{providers} (i.e., social media platforms), are the most powerful element as they possess all the information and can directly impact other groups. Clearly, providers represent a key element that has to ensure the alignment to the trustworthy aspects of RSs. Among the typical objectives of providers are: maximizing the profit, increasing the dwell-time, or reducing the churn-rate. In other non-commercial applications, these can be, i.e., increasing knowledge, improving health, or recommending most enjoyable route in a museum~\cite{Amato}. Platforms face the most comprehensive regulatory obligations, with the DSA requiring disclosure of RS parameters, user modification options, and non-profiling alternatives~\cite[Article 38]{DSA2022} for very large platforms.

However, from the perspective of trustworthy MSRSs, it is necessary to consider the objectives of other stakeholders affected by the system. We should consider the developers of these systems, whose objectives are mostly connected to the technical properties of the RSs, e.g., recommendation complexity or computation time~\cite{Li2023}. Similarly, regulators and policymakers aim to set the requirements for other stakeholder groups to follow. For instance, in the case of social media platforms, an expandability of main parameters of RS~\cite{Fabbri2023} can be required. Following the regulations and/or trustworthiness of RSs, it is necessary to assess whether the RS meets the required criteria. Thus, auditors, as another stakeholder affected by RSs, focus on the degree of auditability of specific RS~\cite{Srba}. Last but not least, we might consider the objectives of society as a whole, which might include general well-being, democratic values, and rule of law. However, it is important to note that the objectives of society will vary, especially when we include cultural considerations in different parts of the world.

Given that every stakeholder has its own objectives in relation to RSs, various value tensions can arise between them~\cite{stray_building_2024, schedl_technical_2025}. In other words, there are trade-offs not only between different stakeholders and their objectives, but also between the different values. For example, content creators may want to promote specific values on social media platforms, such as a healthy lifestyle, i.e., through dietary videos or sponsored content. Such content may not seem problematic, but when we start to promote it to individuals, such as those who suffer from eating disorder, the stream of such videos may have a negative impact on their well-being~\cite {stray_building_2024}. Another value tension can be found between the platforms obligation stemming (Article 27 of DSA) to ``disclose the main parameters used in their recommender systems'' and the platform objective to protect such attributes of the RSs because of a trade secret.

Several authors~\cite{abdollahpouri_multistakeholder_2020, milano_ethical_2021, fabbri_social_2023, deldjoo_fairness_2024, stray2021optimizingforaligningrecommender}
 have pointed that through MSRSs and MORSs, we may find balance between objectives and values of multiple stakeholders and at the same time enhance our understanding of the broader societal impacts, inc. cultural considerations. We argue that research on trustworthy MSRSs is needed, especially when discussing the positive aspect of trustworthiness, which seeks to emphasize the importance of human values.

\section{Future Directions}
\textbf{Trustworthy MORS.} The computer science literature has been primarily concerned with the question what MSRSs and MORSs are and how to create such systems for particular applications~\cite{ZAIZI2023102233}. However, it has not delved deeper into specific solutions, especially when considering not only the complexity of the stakeholders involved and influenced by them, but also the trustworthy aspects. For instance, the fairness, widely researched in standard recommenders, needs a special attention in the MORS setting (e.g., which stakeholders should be treated fair).

\noindent\textbf{Human values.} Another challenge is how to operationalize human values in the design of MSRSs. It is difficult to measure human values as a concrete definition is missing (i.e.,
``well-being'' is contextual and varies among different stakeholders). There are prerequisites (i.e., techniques to measure adherence to values) that are important to consider, but specific methods and solutions are needed to incorporate values of stakeholders at once~\cite{stray_building_2024}. We suggest that case-based and context-sensitive ethical assessments can move beyond abstract ethical principles by including systematic stakeholder identification~\cite{gavornik_utilising_2022}. However, these methods must go beyond identifying stakeholders - they should capture the interplay of stakeholder values, objectives, and potential tension in between.  

\noindent\textbf{Computational complexity.} It is clear, that the multiple objectives coming from different stakeholders can bring overhead from the computation complexity point of view. Especially when a conflicting objectives are on the table, some kind of guidance is required. Incorporating human values can on the one side help to deal with such conflicting objectives. On the contrary, additional objectives are introduced by the trustworthy requirements, increasing the computational complexity of such RSs, demeaning novel technical solutions.

\noindent\textbf{Regulatory requirements.} The complex regulatory environment surrounding RSs complicates the development of trustworthy MRSRs. It requires careful consideration of the legal obligations while balancing technical capabilities and stakeholder interests. For instance the DSA requires the platform to communicate reasons for the relative importance of main parameters used for recommendation, which may be in case of MORS and MSRSs challenging.

\begin{acks}
Partially funded by the NextGenerationEU through the Recovery and Resilience Plan for Slovakia, project No. 09I03-03-V04-00336; and by European Union, project Low Resource Artificial Intelligence, GA No. 101136646.
\end{acks}

\bibliographystyle{ACM-Reference-Format}
\bibliography{sample-base}


\end{document}